\documentclass[aps,prd,twocolumn,showpacs,groupedaddress,nofootinbib]{revtex4}
\usepackage{bm}
\usepackage{amssymb,amsmath,amsthm}

\def\d{{\mathrm{d}}}

\newcommand{\eq}{\begin{equation}}
\newcommand{\eeq}{\end{equation}}
\newcommand{\be}{\begin{equation}}
\newcommand{\ee}{\end{equation}}
\newcommand{\bea}{\begin{eqnarray}}
\newcommand{\nn}{\nonumber}
\newcommand{\eea}{\end{eqnarray}}

\begin{document}
\title{Lower-dimensional Ho\v rava--Lifshitz gravity}
\author{Thomas P. Sotiriou$^1$, Matt Visser$^2$ and Silke Weinfurtner$^3$}
\affiliation{${}^1$Department of Applied Mathematics and Theoretical Physics,
Centre for Mathematical Sciences,
University of Cambridge, Wilberforce Road, Cambridge CB3 0WA, UK\\
${}^2$School of Mathematics, Statistics, and Operations Research,
Victoria University of Wellington, PO Box 600, Wellington 6140, New Zealand\\
${}^3$SISSA - International School for Advanced Studies, 
Via Bonomea 265, 34136, Trieste, Italy
{\rm and} INFN, Sezione di Trieste}

\begin{abstract}
We consider Ho\v rava--Lifshitz gravity in both $1+1$ and $2+1$ dimensions. 
These lower-dimensional versions of Ho\v rava--Lifshitz gravity are simple enough to be explicitly tractable, but still complex enough to be interesting.
We write the most general (non-projectable) action for each case and discuss the resulting dynamics. In the $1+1$ case we utilize the equivalence with 2-dimensional Einstein-aether theory to argue that, even though non-trivial, the theory does not have any local degrees of freedom. In the $2+1$ case we show that the only dynamical degree of freedom is a scalar, which qualitatively has the same dynamical behaviour as the scalar mode in (non-projectable) Ho\v rava--Lifshitz gravity in $3+1$ dimensions. We discuss the suitability of these lower-dimensional theories as simpler playgrounds that could help us gain insight into the $3+1$ theory. As special cases we also discuss the projectable limit of these theories.  Finally, we present an algorithm that extends the equivalence with (higher order) Einstein-aether theory to full Ho\v  rava--Lifshitz gravity (instead of just the low energy limit), and we use this extension to comment on the apparent naturalness of the covariant formulation of the latter.
\end{abstract}                                           
\date{\today}
\maketitle

\section{Introduction}

Two years ago Ho\v rava proposed a gravity theory (now commonly referred to as Ho\v rava--Lifshitz gravity or simply Ho\v rava gravity) that has some realistic hope to be a UV completion of general relativity \cite{Horava:2009uw}. This is to be achieved by adding to the gravitational action higher-order spatial derivatives without adding higher-order time derivatives. This procedure can lead to a modification of the graviton propagator which renders the theory power-counting renormalizable, without increasing the number of time derivatives in the field equations \cite{Horava:2009uw, power-count1, power-count2}.

Clearly such a theory cannot treat space and time on the same footing, and is naturally constructed in terms of a preferred foliation. This foliation can be described by a scalar field, which is an extra degree of freedom of the theory with respect to general relativity. The theory is not invariant under the full set of diffeomorphisms, but it can still be invariant under the more restricted foliation-preserving diffeomorphisms, $t\to\tilde{t}(t)$, $x^i\to\tilde{x}^i(t,x^i)$. It is convenient to consider the Arnowitt--Deser--Misner decomposition of spacetime
\begin{equation}
\d s^2 = - N^2 c^2 \d t^2 + g_{ij}(\d x^i + N^i \d t) (\d x^j + N^j \d t).
\end{equation}
Defining the extrinsic curvature as
\begin{equation}
K_{ij} = {1\over2N} \left\{  \dot g_{ij} - \nabla_i N_j - \nabla_j N_i \right\},
\end{equation}
the action of the theory is of the form
\begin{equation}
\label{action}
S=\frac{M_{\rm pl}^2}{2}\int \d^dx \,\d t \, N \sqrt{g} \left\{ K^{ij} K_{ij} - \lambda K^2 +{\cal V}\right\}\, ,
\end{equation}
where Latin indices run from $1$ to $3$, $M_{\rm pl}$ is the Planck mass, $g$ is the determinant of the spatial metric $g_{ij}$, while $\lambda$ is a dimensionless running coupling, and ${\cal V}$ is the part of the Lagrangian that does not contain any time derivatives. In fact, invariance under foliation-preserving diffeomorphisms requires that ${\cal V}$ does not include any terms containing the shift $N^i$, but is instead constructed only with the lapse $N$, the induced metric $g_{ij}$, and their spatial derivatives.

 Power counting renormalizability, on the other hand, requires that ${\cal V}$ include terms with up to $2z$ derivatives, where $z\geq d$, and $d$ is the number of spatial dimensions \cite{Horava:2009uw, power-count1, power-count2}. When $d=3$, one should have at least $z=3$, {\em i.e.}~6th-order,  operators in ${\cal V}$. Additionally, radiative corrections are expected to generate all possible terms compatible with the symmetries of the theory. This leads to a very large number of terms that one needs to include in $\cal V$. Restricting the lapse to be a function of time only, $N=N(t)$, as has been suggested in reference~\cite{Horava:2009uw} in order to match the reduced symmetry of the theory, leads to a much smaller and tractable number of terms \cite{Sotiriou:2009bx, Sotiriou:2009xxx} --- the so-called projectable version of the theory, see references \cite{Weinfurtner:2010hz,Mukohyama:2010xz} for reviews. However, in this version of the theory, as well as in other versions with different restrictions, such as detailed balance \cite{Horava:2009uw}, the extra scalar degree of freedom exhibits undesirable dynamical behaviour, such as instabilities, over-constrained evolution, and strong coupling  at low energies \cite{Sotiriou:2009xxx,bunch1, bunch2, bunch3, bunch4, Kobakhidze:2009zr,bunch5,bunch6,bunch7,bunch8,bunch9,Wang:2010uga}. (We will not consider in this article  models where the action has been modified in order to respect extra symmetries, such as the model with an extra local $U(1)$ symmetry proposed in reference~\cite{Horava:2010zj}. See reference \cite{Sotiriou:2010wn} for a brief review including all of the various versions of Ho\v rava--Lifshitz gravity.)
 
It turns out that the scalar mode is much better behaved if no projectability or detailed balance restriction is imposed \cite{Blas:2009qj}.  Then ${\cal V}$ has the general form
\be
\label{genV}
{\cal V}=\xi \, R+\eta \,a^i a_i+\frac{1}{M_\star^2}L_4+\frac{1}{M_\star^4}L_6\,,
\ee
where $R$ is the Ricci scalar of $g_{ij}$, 
\be
a_i=\partial_i \ln N\,,
\ee
while  $\xi$ and $\eta$ are dimensionless couplings, $M_\star$ is a new mass scale, and $L_4$ and $L_6$ include all possible terms of 4th and 6th order in spatial derivatives respectively. A cosmological constant can also be added to ${\cal V}$ but we neglect it here for simplicity as it is not important for our discussion. The presence of the $a^i a_i$ term, in addition to the standard $R$ term, is enough to alleviate the aforementioned problems at low energies. 
Strong coupling persists even in this version of the theory and, though it is pushed up to high energies, it still constitutes a potential threat for UV completeness \cite{Papazoglou:2009fj,Kimpton:2010xi}. However, it can be altogether avoided by assigning a specific hierarchy between the scales $M_{pl}$ and $M_\star$ \cite{Blas:2009ck}. 

There is still a long way to go before one can say with confidence whether Ho\v rava's proposal is a truly viable UV complete gravity theory. Renormalizability beyond power-counting has not been demonstrated, and the  renormalization group flow of the various running couplings is not yet known (despite the fact that infrared  viability hinges on the hope that various parameters will run sufficiently rapidly to desired values). Furthermore, there are  various phenomenological aspects of the theory which have not yet been studied, which will lead to new constraints ({\em e.g.}~coupling to matter, equivalence principle violations, etc.).  Despite these limitations, for the moment Ho\v rava--Lifshitz gravity definitely seems to be an interesting candidate for a UV complete gravity theory, one  which deserves further study. 

A purely technical difficulty of the theory is the very large number of terms that one needs to consider as part of $L_4$ and $L_6$. One way to deal with this problem without imposing restriction to the action would be to study the theory in less that 3 spatial dimensions. Since renormalizability requires $z=d$ (at least), taking $d<3$ would also reduce the number of spatial derivatives one would have to allow, so it would also consequently reduce drastically the number of higher-order operators. Therefore, one could consider studying lower-dimension Ho\v rava--Lifshitz gravity in an attempt to gain insight into the $3+1$ dimension theory.

As discussed, the problem with imposing restrictions to the action in order to reduce the number of higher-order operators is essentially that it alters the dynamics of the theory. There is no particular reason to expect that this is not also going to be the case when one reduces the number of spatial dimensions. Indeed, we already know for a fact that the dynamics of general relativity changes drastically when the number of spatial dimensions drops below $3$. Our goal here is precisely to examine to which extent the dynamics of lower-dimensional Ho\v rava--Lifshitz gravity resembles that of the $3+1$ dimensional theory. This will allow us to gauge how much we can learn for the latter from the former. In what follows, we will separately consider $1+1$ and $2+1$ Ho\v rava--Lifshitz gravity, we will construct the full actions and discuss their dynamics, utilizing (and extending to lower dimensions) also the equivalence \cite{Jacobson:2010mx} between Einstein-aether theory \cite{Jacobson:2000xp,Jacobson:2008aj} and the low energy limit of Ho\v rava--Lifshitz gravity.

Note that lower-dimensional models of Ho\v rava--Lifshitz gravity are also interesting in their own right. Renormalizable gravity theories, even in less that $3+1$ dimensions, are not so easy to find. Such models could potentially be used as duals to non-relativistic quantum field theories in the context of the AdS/CFT correspondence, or as theories describing gravity on worldsheets of strings or worldvolumes of branes \cite{Horava:2008ih}. See also reference~\cite{Horava:2011gd} for further examples.

\section{Ho\v rava--Lifshitz gravity in $1+1$ dimensions}
\subsection{Setup and most general action}

We start by considering the simplest case of $1+1$ dimensions. The action of the theory in this case is drastically simplified as 1-dimensional space cannot have intrinsic curvature. At the same time the extrinsic curvature is actually a scalar, given by
\be
K= {1\over2N} \left\{  \dot g_{11} - 2\nabla_1 N_1\right\},
\ee
Finally, we have $z=d=1$, so ${\cal V}$ needs to contain space-covariant terms we can construct with $g_{ij}\to g_{11}$, and $a_i\to a_1$, with up to 2 spatial derivatives only. Thus, the action has the simple form
\begin{equation}
\label{1daction}
S=\frac{M_{\rm pl}^2}{2}\int \d x \, \d t \, N \sqrt{g_{11}} \left\{ (1 - \lambda) \; K^2 +\eta \;g^{11}a_1a_1\right\}\, .
\end{equation}
When $\lambda=1$ and $\eta=0$, which corresponds to the values of these parameters in general relativity, the theory becomes trivial as expected. We will not attempt to derive field equations for this theory. Instead, in the next section we will establish its dynamical equivalence with Einstein-aether theory in 2 dimensions, by applying the results of reference~\cite{Jacobson:2010mx}. We will then use this equivalence to discuss the dynamics.

\subsection{Equivalence with Einstein-aether theory and dynamics}
\label{1+1equiv}

The action for Einstein-aether theory \cite{Jacobson:2000xp,Jacobson:2008aj} is 
\be \label{S}
S^{4d}_{\ae} = \frac{1}{16\pi G_{\ae}}\int \d^{4}x\, \sqrt{-\tilde{g}}~ (-\tilde{R} -M^{\alpha\beta\mu\nu} \tilde{\nabla}_\alpha u_\mu \tilde{\nabla}_\beta u_\nu)
\ee
where Greek indices run from $0$ to $3$, $\tilde{R}$ is the 4-dimensional Ricci scalar of the spacetime metric $\tilde{g}_{ab}$, $\tilde{g}$ is the determinant of that metric, $ \tilde{\nabla}_\alpha$ the associated covariant derivative, and
\be
M^{\alpha\beta\mu\nu} = c_1 \tilde{g}^{\alpha\beta}\tilde{g}^{\mu\nu}+c_2\tilde{g}^{\alpha\mu}\tilde{g}^{\beta\nu}+c_3 \tilde{g}^{\alpha\nu}\tilde{g}^{\beta\mu}+c_4 u^\alpha u^\beta \tilde{g}_{\mu\nu}\,, 
\ee
(a tilde will be used to denote a spacetime metric, as opposed to an induced spatial metric, irrespectively of dimensionality).  $G_{\ae}$ has dimensions of an inverse mass squared, whereas the parameters $c_i$ are dimensionless. Additionally, the aether field, $u_\mu$ is forced to satisfy the constraint $u_\mu u^\mu=1$. Here we assume that this constraint is imposed implicitly by allowing only variations that respect it. Alternatively, one could impose it explicitly, by the use of a Lagrange multiplier. It has been shown in reference~\cite{Jacobson:2010mx} that, once the extra restriction that the aether is hypersurface orthogonal has been imposed, this aether action is dynamically equivalent to the low-energy limit of Ho\v rava--Lifshitz gravity in $3+1$ dimensions, {\em i.e.}~to the action given in equation~(\ref{action}), \emph{but without the higher-order operators $L_4$ and $L_6$}. Given also the unit vector constraint on the aether, this restriction amounts locally to the requirement that there exists a function such that
\be
\label{ho}
u_\alpha=\frac{\partial_\alpha T}{\sqrt{\tilde{g}^{\mu\nu}\; \partial_\mu T \partial_\nu T}}\,,
\ee 
However, one could choose to work in a gauge where $T$ is identified with the time coordinate $t$. Then
\be
\label{hou}
u_\mu=\delta_{\mu T}\; (g^{TT})^{-1/2}=N\delta_{\mu T}\,.
\ee
If one uses equation~(\ref{hou}) to replace $u_\mu$ in equation~(\ref{S}), the low-energy limit of the action of Ho\v rava--Lifshitz gravity in $3+1$ dimensions is recovered, that is, the action given in equation~(\ref{action}) without the higher-order operators $L_4$ and $L_6$, and with the following correspondence of parameters:
\bea
\label{HLpar}
&&\quad\frac{1}{8\pi M_{pl}^2 \,G_{\ae}}=\xi=\frac{1}{1-c_{13}},\\\nn\\&& \lambda=\frac{1+c_2}{1-c_{13}},\quad\qquad \eta=\frac{c_{14}}{1-c_{13}},\nn
\eea
where $c_{ij}=c_i+c_j$.

Clearly, this equivalence will apply also apply in 2 dimensions (or indeed any number of dimensions). The action of 2-dimensional Einstein-aether theory is \cite{Eling:2006xg}
\be \label{S2}
S^{2d}_{\ae} = -\frac{1}{16\pi G_{\ae}}\int \d^{2}x\; \sqrt{-\tilde{g}}L^{2d}_{\ae} ,
\ee
where
\be
L^{2d}_{\ae}=\frac{1}{2}c_{14} F^{\alpha\beta}F_{\alpha\beta}+c_{123}(\tilde{\nabla}_\alpha u^\alpha)^2\,,
\ee
and we have ignored total divergences (such as the $\tilde{R}$ term). Greek indices now take the values $0$ and $1$ only, and
\be
F_{\alpha\beta}=\tilde{\nabla}_\alpha u_\beta-\tilde{\nabla}_\beta u_\alpha\,.
\ee
In 2 dimensions any vector field is always locally hypersurface orthogonal and, thus, so is the aether. It is straightforward to show that, after choosing $T$ as the time coordinate,
\bea
&&\tilde{\nabla}_\alpha u^\alpha=-K\,,\qquad F^{\alpha \beta}F_{\alpha\beta}=-g^{11}a_1 a_1\,.
\eea
Then, the action in equation~(\ref{S2}) becomes the action of Ho\v rava--Lifshitz gravity in $1+1$ dimensions, as given in equation~(\ref{1daction}), with the parameter correspondence as given in equation~(\ref{HLpar}).

As was mentioned earlier, 4-dimensional Einstein-aether theory with the extra restriction of the aether being hypersurface orthogonal is equivalent to the low-energy limit of  Ho\v rava--Lifshitz gravity in $3+1$ dimensions, not the full theory. However, in Ho\v rava--Lifshitz gravity in $1+1$ dimensions no operators with more than two spatial derivatives need to be considered anyway, as naive power counting renormalizability demands $z=d=1$. Therefore, the complete $1+1$ Ho\v rava theory  turns out to be equivalent to 2-dimensional Einstein-aether theory. Additionally, it is worth stressing once more that in 2 dimensions the aether is necessarily hypersurface orthogonal, so no further restriction on the Einstein-aether side is needed either.

In reference~\cite{Eling:2006xg} the full set of solutions of $2$-dimensional Einstein-aether theory has been found. Though nontrivial, unlike general relativity in 2 dimensions, the theory does not posses any local degrees of freedom. The equivalence presented above can be used to turn these solutions of Einstein-aether theory into solutions of Ho\v rava--Lifshitz gravity in $1+1$ dimensions --- then similar conclusions can be made for the dynamics of the latter. Based on this, although it might constitute an interesting model for a 2-dimensional quantum gravity theory, we see that Ho\v rava--Lifshitz gravity in $1+1$ dimensions is drastically different from its $3+1$ counterpart. 

 It is also worth mentioning that naive  power-counting renormalizability should be taken with a grain of salt in $1+1$ dimensions. Since there are no local degrees of freedom the perturbative power-counting arguments do not (strictly speaking) apply.  However, again due to this lack of local degrees of freedom,  the theory is likely to still be non-perturbatively renormalizable, like 2+1 general relativity.
 
\subsection{$1+1$ projectable theory as a limiting case}

\label{proj1+1}

As mentioned in the Introduction, our main interest  is the most general  Ho\v{r}ava--Lifshitz gravity without any restrictions, as the more restricted versions tend to have viability and consistency problems.  However, it is rather trivial for one to obtain the most general action and understand the dynamics of projectable Ho\v rava--Lifshitz gravity as a limiting case of what has been presented above. The difference between the projectable version and the version we considered here is that in the former the lapse is forced to be a function of time only, {\em i.e}~$N=N(t)$, at the level of the action.

In $1+1$ dimensions requiring $N=N(t)$ would make the last term in eq.~(\ref{1daction}) vanish, so the most general action in the projectable version is
\begin{equation}
\label{1dactionp}
S_p=\frac{M_{\rm pl}^2}{2}\int \d x \, \d t \, N \sqrt{g_{11}} (1 - \lambda) \; K^2 \, .
\end{equation}
Clearly, provided that $\lambda\neq 1$ and possibly modulo a sign, the factor $(1-\lambda)$ can be absorbed in a redefinition of $M_{\rm pl}$. So, the crucial difference between projectable Ho\v rava--Lifshitz gravity and general relativity in $1+1$ dimensions is simply the fact that $N=N(t)$, and nothing more.

\section{Ho\v rava--Lifshitz gravity in $2+1$ dimensions}
\subsection{Setup and most general action}

We now turn our attention to 2 spatial dimensions. The action can take the form given in equation~(\ref{action}), with $K_{ij}$ being the extrinsic curvature of the 2-dimensional spatial hypersurfaces. What remains is to determine ${\cal V}$. We have $z=d=2$, so ${\cal V}$ needs to contain all space-covariant terms we can construct with $g_{ij}$ and $a_i$, with up to 4 spatial derivatives. Some of these terms, however, can be eliminated if one takes into account the following:
\begin{itemize}
\item
In 2 dimensions we know that
\be
R_{abcd}=\frac{1}{2}(g_{ac}g_{db}-g_{ad}g_{cb}) R,
\ee
so all curvature invariants can be expressed in terms of the Ricci scalar.
\item Various terms differ only by a total divergence.
\item The vector $a_i$ is the gradient of a scalar.
\end{itemize}
Once all of the above have been taken into account one can, without any loss of generality, write
\bea
{\cal V}&=&\xi\,R+\eta\, a^i a_i+g_1 \,R^2+g_2\, \nabla^2R+g_3\,(a^ia_i)^2\nn\\
&&+g_4\, Ra^ia_i+g_5 a^2 (\nabla\cdot a) + g_6 (\nabla\cdot a)^2 \nn\\
&&+ g_7 (\nabla_i a_j) (\nabla^i a^j).
\eea
The $g_i$ couplings are not dimensionless, but have dimensions of an inverse mass squared ({\em i.e.}~we have absorbed the mass scale $M_\star$ into these couplings). Again, we can add a cosmological constant which we neglect here for simplicity. The most general action in $2+1$ dimension then has the form
\bea
\label{2daction}
S&=&\frac{M_{pl}^2}{2}\int \d^2 x \, \d t \, N\sqrt{g} \Big\{K^{ij}K_{ij}-\lambda K^2+\xi R +\eta\, a_i a^i \nn\\
&&\qquad\qquad+g_1 \,R^2+g_2\, \nabla^2R+g_3\,(a^ia_i)^2\nn\\&&\qquad\qquad+g_4\, Ra^ia_i+g_5 a^2 (\nabla\cdot a) + g_6 (\nabla\cdot a)^2 \nn\\&&\qquad\qquad+ g_7 (\nabla_i a_j) (\nabla^i a^j)\Big\}.
\eea
We can obtain the field equations by varying the action with respect to the lapse $N$, the shift $N^i$, and the induced metric $g_{ij}$.
Variation with respect to the lapse yields
\bea
\label{eqlapse}
&&K^{ij}K_{ij}-\lambda K^2+\xi R+ \eta\left[2 \nabla\cdot a +a^ia_i\right]\\&&-g_1 R^2-g_2 \nabla^2 R+g_3\left[4\nabla_i(a_ja^j a^i)+3(a_ia^i)^2\right]\nn\\&&+g_4 \left[2 \nabla_i(R a^i) +Ra^ia_i\right]+g_5\big\{a^ia_i \nabla\cdot a\nn\\&&+2\nabla\cdot (a[\nabla\cdot a])-[\nabla^2(Na^ia_i)]/N\big\}\nn\\&&-g_6\big[3 (\nabla\cdot a)^2 + 2 \nabla^2 (\nabla\cdot a) + 2 a^2 (\nabla\cdot a) \nn\\&&+4 (a\cdot\nabla) (\nabla\cdot a)\big] 
-g_7\big[ (\nabla_i a_j) (\nabla^i a^j) \nn\\&&-2 (\nabla_i a_j) a^i a^j -4 \nabla^i ( a^j (\nabla_i a_j) ) - 2 \nabla_i \nabla^2  a^i\big]\,
=0.\nn\\\nn
\eea

Variation with respect to the shift yields
\be
\label{eqshift}
\nabla_i\pi^{ij}\equiv \nabla_i\left\{K^{ij}-\lambda Kg^{ij}\right\}=0\,.
\ee
\def\L{{\mathcal{L}}}
Variation with respect the spatial metric $g_{ij}$ yields
\begin{widetext}
\begin{eqnarray}
\label{piequation}
&&\frac{1}{\sqrt{g}} \; \partial_t ( \sqrt{g} \; \pi_{ij} ) +2N \left(K_i^{\phantom{a}l}K_{lj} - \lambda K K_{ij}\right) 
- {1\over2} N \left(K^{ij}K_{ij}-\lambda K^2\right) g^{ij} 
- N  \pi^{ij} (\nabla_m N^m)- \L_{\vec N} (N\pi_{ij}) 
\nonumber\\
&& +2 N N_{(i}\pi_{j)m}a^m-\xi \frac{1}{N}[\nabla_i\nabla_j -g_{ij}\nabla^2]N +\eta\,\left[a_i a_j -\frac{1}{2} a^l a_lg_{ij}\right]+\frac{1}{2} g_1R^2 g_{ij} -2 g_1\frac{1}{N}[\nabla_i\nabla_j -g_{ij}\nabla^2] (NR)\nn\\&&-g_2 \frac{1}{N}[\nabla_i\nabla_j -g_{ij}\nabla^2] (\nabla^2 N)  -g_2 a_{(i} \nabla_{j)} R 
+g_2 {\nabla_m(NRa^m)\over2N}  g_{ij}+2 g_3 a^2 a_i a_j -{1\over2} g_3(a^2)^2 g_{ij}+g_4
R a_i a_j\nn\\&&
-g_4 \frac{1}{N}[\nabla_i\nabla_j -g_{ij}\nabla^2] (N a^i a_i)  - {1\over2}  g_4 Ra^2 g_{ij} +g_5  a_i a_j [(\nabla\cdot a) - a^2] 
-{1\over2}g_5  \left\{ (\nabla_i a^2) a_j + (\nabla_j a^2) a_i \right\}\nn\\&&
 + {1\over2} g_5\left\{ (a^2)^2+ a^m\nabla_m a^2 \right\} g_{ij}-2g_6 (\nabla\cdot a) a_i a_j  -g_6\left\{ (\nabla_i [\nabla\cdot a]) a_j + (\nabla_j i [\nabla\cdot a) a_i \right\}\nn\\
&& 
 +  g_6\left[a^2  (\nabla\cdot a) +  a\cdot\nabla (\nabla\cdot a) +{1\over2}  (\nabla\cdot a)^2 \right]g_{ij}
 -{1\over2} g_7 (\nabla_m a_n) (\nabla^m a^n)  g_{ij} - 2 g_7(a\cdot\nabla a_{(i}) a_{j)} -2 g_7(\nabla^2 a_{(i}) a_{j)}\nn\\
&&
+
g_7 a^2 (\nabla_i a_j) + g_7\nabla\cdot( [\nabla_i a_j] a)=0\,,
\eea
\end{widetext}
where $ \L_{\vec N}$ denotes the Lie derivative along the vector $N^i$.

Though rather lengthy, these equations are perhaps more manageable than they seem. Equations~(\ref{eqlapse}) and (\ref{eqshift}) do not contain any time derivatives of the lapse or the shift and are, therefore, constraints. The only dynamical equation is~(\ref{piequation}) and the only dynamical variable is $g_{ij}$. However, applying the uniformization theorem
\cite{poincare}, since $g_{ij}$ is a 2-dimensional metric and the theory is invariant under transformation of the sort $x^i\to\tilde{x}^i(t,x^i)$, there is enough gauge freedom to set
\be
\label{gauge}
g_{ij}=\Omega^2\;  g^c_{ij}\,,
\ee
where $g^c_{ij}$ denotes the metric of a constant curvature spherical, Euclidean or hyperbolic 2-dimensional space. This gauge choice would turn equation~(\ref{piequation}) into a dynamical equation for the conformal factor $\Omega$. We will not proceed further to explicitly write the equation in this specific gauge, as the current discussion suffices to convincingly argue that the theory propagates only a single scalar degree of freedom. 

\subsection{Linearization and dynamics}

To provide further support for our claim that there is only a single scalar degree of freedom, we linearize the theory around flat 2+1 Minkowski space. This will also allow us to determine the linearized propagator for the scalar, and get a deeper insight into the dynamics. We start with the action given in equation~(\ref{2daction}),  and we perturb to quadratic order. We have 
\be
N=1+\alpha, \qquad\qquad N_i= 0 + n_i, 
\ee
and we make the gauge choice
\be
g_{ij}=e^{2\zeta} \delta_{ij}.
\ee
The terms $(a^ia_i)^2$, $Ra^ia_i$ and $a^2 (\nabla\cdot a)$ will not contribute to this order (since by construction $a_i=R=0$ in the background). On the other hand, $(\nabla\cdot a)^2$ and $(\nabla_i a_j) (\nabla^i a^j)$ give the same (non-trivial) contribution to this order, so we define (with some foresight) $g_{67}=(g_6+g_7)/2$. The quantity $K_{ij}$ appears only quadratically in the action, so we only need to compute it to first order:
\bea
&&K^{(1)}_{ij}=\dot{\zeta} \delta_{ij}-\partial_{(i} n_{j)}\,,\\
&&K^{(1)}= 2 \dot{\zeta} -\partial_i n^i \,.
\eea
The  quadratic action then takes the form
\bea
S&=&M_{pl}^2\int \d^2 x \, \d t \, \Big\{(1-2\lambda)\dot{\zeta}^2- (1-2\lambda)\dot{\zeta} (\partial_i n^i)\nn\\
&&\quad\qquad+\frac{1}{4}(1-2\lambda) (\partial_i n^i)^2
+{1\over4} n^i \partial^2 n_i   -\xi \alpha \partial^2 \zeta \nn\\
&&\quad\qquad+\frac{\eta}{2} (\partial_i\alpha)(\partial^i\alpha) 
+2 g_1(\partial^2\zeta)^2\nn\\
&&\quad\qquad-2 g_2 (\partial^2 \alpha) (\partial^2 \zeta)+g_{67}(\partial^2 \alpha)^2\Big\},
\eea
where $\partial^2=\delta^{ij}\partial_i\partial_j$. 

We start by varying with respect to $n_i$. This variation yields
\be
\label{neq}
{1\over2}(1-2\lambda)\partial_i (\partial_j n^j) +{1\over2} \partial^2 n_i =(1-2\lambda) \partial_i \dot{\zeta}\,.
\ee
Taking the divergence leads to the equation
\be
\label{interimni}
(1-\lambda)\partial^2 (\partial_i n^i) =(1-2\lambda) \partial^2 \dot{\zeta}\,,
\ee
which, (given suitable regular boundary conditions), can be trivially integrated to give
\be
\label{interimbeta}
\partial_i n^i=\frac{1-2\lambda}{1-\lambda} \dot{\zeta}\,.
\ee
Re-inserting this into equation~(\ref{neq}) one gets
\be
\partial_i (\partial_j n^j) -\partial^2 n_i =0\,.
\ee
In particular, this implies
\be
\partial^2 n_{[i,j]} =0\,,
\ee
which in turn (given suitable regular boundary conditions) implies that 
\be
\label{eqni}
n_{[i,j]} =0 \quad \Rightarrow \quad n_i =\partial_i \beta\,.
\ee
That is, $n_i$ has to be the gradient of a scalar, which leaves no room for vector perturbations. From equation~(\ref{interimbeta}) we get
\be
\label{eqbeta}
\partial^2\beta=\frac{1-2\lambda}{1-\lambda} \; \dot{\zeta}\,.
\ee

We now move on to the variation with respect to $\alpha$. This  yields
\be
-g_2\partial^4 \zeta+2g_{67} \,\partial^4 \alpha-\xi \,\partial^2\zeta -\eta \,\partial^2\alpha=0.
\ee
Again, imposing some regularity conditions, this can be solved to give
\be
\label{eqalpha}
\alpha=-\frac{\xi +g_2 \partial^2}{\eta-2 g_{67}\partial^2} \; \zeta\,.
\ee
We can now use equations~(\ref{eqni}),  (\ref{eqbeta}) and (\ref{eqalpha}) to integrate out the non-dynamical fields $n_i$ and $\alpha$ in favour of the dynamical field $\zeta$. The quadratic action then reads
\bea
S=M_{pl}^2\int \d^2 x \,\d t \,\Big\{\frac{1}{2} \; \frac{1-2\lambda}{1-\lambda}\; \dot{\zeta}^2+\zeta D \zeta\Big\}\,,
\eea
where
\be
D\equiv \frac{\xi^2+2\left[2\eta g_1+\xi g_2\right]\partial^2+\left[g_2^2-4 g_1(g_6+g_7)\right]\partial^4}{2(\eta- (g_6+g_7)\partial^2)} \; \partial^2\,.
\ee
The dispersion relation for the scalar is then given by a rational polynomial
\be
\label{dispersion}
\frac{1}{2}\;\frac{1-2\lambda}{1-\lambda} \; \omega^2= 
\frac{P_1(k)}{P_2(k)} \,,
\ee
where
\bea
P_1(k)&=&\xi^2 k^2-2\left[2\eta g_1+\xi g_2\right]k^4\nn\\&&+\left[g_2^2-4 g_1 (g_{6}+g_{7})\right]k^6\\
P_2(k)&=&2(\eta+(g_{6}+g_{7}) k^2)\,.
\eea
Note that this rational polynomial dispersion relation is qualitatively similar to that of the scalar mode in $3+1$ dimensions \cite{Blas:2009qj}. Classical stability requires this action to have the correct relative sign between the kinetic and the potential term, that is, it requires
\be
\frac{1-\lambda}{1-2\lambda}\;\frac{1}{\eta}>0\,.
\ee
Since we have only one excitation, positivity of energy is not really a concern, as it can be controlled by flipping the overall sign of the action. This is not the case in $3+1$ dimensions, where there is a spin-2 graviton as well, and one needs to separately require that its kinetic term has the same sign as the kinetic term of the scalar, if neither of the two fields is to be a ghost.

In the low-momentum limit we have
\be
\label{eqalpha2}
\alpha\approx -\frac{\xi}{\eta} \; \zeta\,, \qquad D \approx \frac{\xi^2}{2\eta} \; \partial^2\,.
\ee
That is, $\alpha$ is given algebraically in terms of the conformal factor of the spatial metric $\zeta$, which now satisfies a standard linear dispersion relation
\be
\label{dispersion2}
\frac{1}{2}\;\frac{1-2\lambda}{1-\lambda} \; \omega^2 \approx
\frac{\xi^2 k^2}{2\eta}\,,
\ee
with (low momentum) phase velocity
\be
c_\zeta=\frac{\xi}{\sqrt{\eta}}\sqrt{\frac{1-\lambda}{1-2\lambda}}\,.
\ee
Clearly, the limit to general relativity is far from trivial. In that limit we have $g_i\to 0$, $\eta\to 0$ and $\lambda\to 1$, so the behaviour of the mode depends strongly on how $\eta/(\lambda-1)$ scales as both numerator and denominator approach zero. Note that in general relativity, where $g_i=0$, $\eta= 0$ and $\lambda= 1$ {\em a priori}, equation~({\ref{interimni}) is enough to render the dynamics of $\zeta$ trivial.

What about the dynamics beyond the quadratic order? One could straightforwardly follow the lines of reference~\cite{Papazoglou:2009fj} (or reference~\cite{Kimpton:2010xi}) and derive the cubic interactions for the scalar. However, we will not attempt this here, simply because the calculation is not sensitive to the dimensionality of space. Thus, there is no reason to believe that the result will differ in any way (at least qualitatively) from that obtained in the case of $3+1$ dimensions. It is, therefore, expected that the scalar will exhibit strong coupling at some scale determined by the magnitudes of $\lambda$ and $\eta$ \cite{Papazoglou:2009fj,Kimpton:2010xi}. It is also expected that choosing the right hierarchy between $M_{pl}$ and $M_\star$ will alleviate this strong coupling, just as in the $3+1$ case \cite{Blas:2009ck}. The only difference here is that, since in 2+1 dimensions there are no experimental constraints but only consistency requirements, one now has the freedom to choose the magnitude of $\lambda$ and $\eta$ so as to push the strong coupling scale beyond the energies at which the higher-order operators become important.

To sum up, the theory in $2+1$ dimensions clearly has a fundamental difference from its counterpart in $3+1$ dimensions, namely the absence of a spin-2 graviton, as one might have expected. However, it does possess a scalar degree of freedom with non-trivial dynamics. Additionally, this degree of freedom exhibits (qualitatively) the same dynamical behaviour as  does the scalar degree of freedom in $3+1$ dimensional Ho\v rava--Lifshitz gravity. Therefore, it seems that Ho\v rava--Lifshitz gravity in $2+1$ dimensions is a good theoretical playground for exploring scalar-mode related aspects of Ho\v rava--Lifshitz gravity in $3+1$ dimensions.

\subsection{Equivalence with Einstein-aether theory and covariant formulation}

As one can readily see from the discussion in section \ref{1+1equiv}  the equivalence between  Ho\v rava--Lifshitz gravity and Einstein-aether theory with a hypersurface orthogonal aether does not really hinge on the dimensionality. In fact, in $2+1$ dimensions it will go through exactly in the same way as in $3+1$ dimensions. However, unlike the special case of $1+1$ dimensions, where the action of Ho\v rava--Lifshitz gravity does not contain any higher-order terms, here this equivalence will be limited to the low-energy limit only, and not hold for the complete action.
Nonetheless, restricted though it may be, this equivalence is suggestive on its own of a suitable extension to all energies. One could follow this series of steps:
\begin{enumerate}
\item 
Construct the most general (fully covariant) extension of Einstein-aether theory with up to four derivatives, imposing the additional restriction that the aether be hypersurface orthogonal.
\item 
Observe that the part of the action that is second-order in derivatives is equivalent to the low-energy limit of Ho\v rava--Lifshitz gravity, and that if there is a more general equivalence it will necessarily have to work order by order.
\item 
Use the prescription from the second-order equivalence to identify how to implement the preferred foliation of Ho\v rava--Lifshitz gravity to this higher-order Einstein-aether theory.
\item 
Compare the higher-order terms of the two theories order by order, and identify the parameter matching that will lead to the desired equivalence.
\end{enumerate}
The number of covariant terms which are 4th-order in derivatives in 3-dimensional Einstein-aether theory is quite large and, therefore, we will not attempt to apply the algorithm described above and present the result explicitly (doing so at this stage seems also to be of little practical significance). However, as a point of principle, the existence of this algorithm provides a method of ``covariantization'', alternative to the use of projection operators as suggested in reference~\cite{Germani:2009yt}. More importantly, it allows one to make the following interesting observations.

The number of 4th-order terms in equation~(\ref{2daction}) is far smaller than the number of covariant terms which are 4th-order in derivatives in 3-dimensional Einstein-aether theory (even after hypersurface orthogonality of the aether has been imposed). Indeed, the latter generically contains higher-order time derivatives, whereas the former is carefully constructed not to. This will be reflected in the matching of parameters in step 4: several of the couplings of the 4th-order terms in Einstein-aether theory will have to be exactly tuned to satisfy specific algebraic relations in order for the higher-order time derivatives to cancel out. This will reduce the number of independent couplings to exactly the number of independent couplings of the corresponding Ho\v rava--Lifshitz gravity. Though this seems perfectly feasible, it implies that, reformulated as a covariant theory, Ho\v rava--Lifshitz gravity would seem to be unnaturally fine-tuned, even though this does  not really seem to be the case, (at least not according to power-counting renormalizability arguments).

Viewed from a perspective of the corresponding Einstein-aether theory the same observations sound perhaps more striking: When constructing such a theory with terms up to a certain order in derivatives, selectively neglecting to exclude a term, or relating its coefficient with another term's coefficient, would certainly be considered unnatural fine tuning. (Based on this observation, standard Einstein-aether theory includes all possible terms which are second-order in derivatives.) However, the specific fine tuning that leads to the corresponding Ho\v rava--Lifshitz gravity is \emph{not} actually unnatural. Remarkably this is exactly the choice that removes higher-order time derivatives, which would otherwise be very worrisome. (Note that it is not the lack of Lorentz symmetry {\em per se} that make this specific fine tuning natural, as any generic choice of parameters would anyway lead to a Lorentz violating theory.)

Clearly, all of the above observations are not particular to 2 spatial dimensions. Therefore, the whole discussion applies unmodified to the phenomenologically more interesting $3+1$ Ho\v rava--Lifshitz gravity.

\subsection{$2+1$ projectable theory as a limiting case}

We now consider the limit to the $2+1$ dimensional projectable version of the theory, same as we did in section \ref{proj1+1} for $1+1$ dimensions.
In $2+1$ dimensions, starting from eq.~(\ref{2daction}) and setting $N=N(t)$ yields
\be
\label{2dactionp}
S_p=\frac{M_{pl}^2}{2}\int \d^2 x \, \d t \, N\sqrt{g} \left[K^{ij}K_{ij}-\lambda K^2+g_1 \,R^2\right].
\ee
Note that, in addition to all terms involving $a_i$ vanishing, the $g_2 \nabla^2 R$ term in eq.~(\ref{2daction}) can also be neglected in the projectable case, as it becomes a boundary term.  Furthermore, in view of the Gauss-Bonnet theorem in 2 space dimensions, the $\xi R$ term is a total divergence which becomes a pure boundary term when $N=N(t)$. Hence, it should also be discarded.

We will refrain from presenting the field equations for the projectable theory here, but they can be straightforwardly derived from eqs.~(\ref{eqlapse}), (\ref{eqshift}) and (\ref{piequation}), by setting $N=N(t)$ and discarding terms proportional to  $g_2$ and $\eta$. Some extra care is needed in the case of the equation for the lapse, which will now have to be turned to into a global, instead of a local constraint.

What is perhaps of more interest is to obtain the linearized dispersion relation for the projectable theory as a limiting case. Note that taking the limit in which $\eta\to 0$, $g_i\to 0$ for $i\geq 2$ would not be correct, as it would not actually enforce the constraint $N=N(t)$. Instead, if we view the projectable model as a limit of the non-projectable model, then the right limit is $\eta\to \infty$ (this would force $\alpha\to 0$ in the linearized theory). The dispersion relation then becomes
\be
\label{dispersionp}
\frac{1}{2}\;\frac{1-2\lambda}{1-\lambda} \; \omega^2= -2 g_1 k^4\,,
\ee
Remarkably, there is no $k^2$ term in this dispersion relation. This can of course be traced back to the fact that in 2 dimensions the Ricci scalar $R$  is actually a total divergence. 
This property is specific to 2 dimensions, making $2+1$ dimensional projectable Ho\v rava--Lifshitz gravity rather special and non-representative. In fact, in $3+1$ dimensions, a $k^2$ term is present in the dispersion relation of both the spin-2 and the scalar mode \cite{Sotiriou:2009xxx}.

\section{Conclusions}

We have studied (non-projectable) Ho\v rava--Lifshitz gravity in $1+1$ and $2+1$ dimensions. The $1+1$ theory includes only second-order operators and is, as might have been expected, dynamically equivalent to 2-dimensional Einstein-aether theory. Given that the dynamics of the latter is well known, we can easily infer that  Ho\v rava--Lifshitz gravity in $1+1$ dimensions, though non-trivial, does not have any local degrees of freedom.   Therefore, its dynamics (or lack thereof) is qualitatively very different from the $3+1$, theory. In fact, in this regard, $1+1$ Ho\v rava--Lifshitz gravity actually shares features of $2+1$ Einstein gravity.

We then moved on to $2+1$ Ho\v rava--Lifshitz gravity could also be considered very different from  the  $3+1$ theory as it does not have any spin-2 mode. It does, however, have a propagating scalar mode, so it is far from being dynamically trivial. Additionally, if viewed from a different perspective, it actually bears remarkable similarity with the $3+1$ theory, as the propagating scalar mode has, qualitatively,  the same dynamical behaviour (rational polynomial dispersion relation, strong coupling, etc.) as the scalar mode in the $3+1$ theory.

Based on the above, one can easily infer that the study of Ho\v rava--Lifshitz gravity in $1+1$ dimensions cannot provide much insight into the properties of the $3+1$ theory. On the other hand, studying Ho\v rava--Lifshitz gravity in $2+1$ dimensions could provide a simpler and tractable setting for studying the properties of the scalar mode in the $3+1$ theory. Additionally, it could help understand properties of the $3+1$ theory which are not strictly (and solely) dependent on the type of the propagating field, {\em e.g.}~renormalization properties. Clearly, both $1+1$ and $2+1$ theories are interesting in their own right as well.

In addition to the above, we also used the $2+1$ theory and its equivalence with 3-dimensional Einstein-aether theory as an example to discuss a possible covariant formulation. We provided an algorithm for writing Ho\v rava--Lifshitz gravity as a higher-order Einstein-aether theory. Starting from this algorithm, we were able to argue that Ho\v rava--Lifshitz gravity would appear to be unnaturally fine-tuned when seen as a covariant theory, even though this does not seem to be the case according to power-counting renormalizability arguments. Put differently, fine tuning a higher-order Einstein-aether theory to eliminate higher-order time derivatives is remarkably not unnatural, provided that the aforementioned renormalizability arguments are robust beyond power counting.

Lastly, we also briefly considered the projectable version of Ho\v rava--Lifshitz gravity, where the lapse function is assumed to be space-indepedent, as a limiting case of the results we had already obtained. In $1+1$ dimensions, the action simply reduces to that of general relativity but with the projectability constraint. In $2+1$ dimensions, the action is that of general relativity with an extra $R^2$ term and the projectability constraint. There is a propagating scalar degree of freedom satisfying a dispersion relation of the form $\omega^2\propto k^4$. The absence of the $k^2$ is particular to $2+1$ dimensions making the $2+1$ theory non-representative (and Lorentz-violating at all scales).

\section*{Acknowledgments}
TPS and SW were supported by Marie Curie Fellowships. MV was supported by the Marsden Fund administered by the Royal Society of New Zealand. We further acknowledge partial support via a FQXi travel grant.




\begin{thebibliography}{99}



\bibitem{Horava:2009uw}
  P.~Ho\v{r}ava,
  ``Quantum Gravity at a Lifshitz Point,''
  Phys.\ Rev.\  D {\bf 79}, 084008 (2009)
  [arXiv:0901.3775 [hep-th]].
  
  \bibitem{power-count1}
  M.~Visser,
  ``Lorentz symmetry breaking as a quantum field theory regulator,''
  Phys.\ Rev.\  {\bf D80 } (2009)  025011.
  [arXiv:0902.0590 [hep-th]].
  
  \bibitem{power-count2}
  M.~Visser,
  ``Power-counting renormalizability of generalized Ho\v{r}ava gravity,''
  [arXiv:0912.4757 [hep-th]].
 
   
\bibitem{Sotiriou:2009bx}
  T.~P.~Sotiriou, M.~Visser and S.~Weinfurtner,
  ``Phenomenologically viable Lorentz-violating quantum gravity'',
  Phys.\ Rev.\ Lett.\  {\bf 102}, 251601 (2009)
  [arXiv:0904.4464 [hep-th]].
  
  
  \bibitem{Sotiriou:2009xxx}
  T.~P.~Sotiriou, M.~Visser and S.~Weinfurtner,
  ``Quantum gravity without Lorentz invariance,''
  JHEP {\bf 0910}, 033 (2009)
  [arXiv:0905.2798 [hep-th]];
  
 
\bibitem{Weinfurtner:2010hz}
  S.~Weinfurtner, T.~P.~Sotiriou and M.~Visser,
  ``Projectable Ho\v{r}ava--Lifshitz gravity in a nutshell,''
  J.\ Phys.\ Conf.\ Ser.\  {\bf 222}, 012054 (2010)
  [arXiv:1002.0308 [gr-qc]].

\bibitem{Mukohyama:2010xz}
  S.~Mukohyama,
  ``Ho\v{r}ava--Lifshitz Cosmology: A Review,''
  Class.\ Quant.\ Grav.\  {\bf 27}, 223101 (2010)
  [arXiv:1007.5199 [hep-th]].

 
  
\bibitem{bunch1}
  C.~Charmousis, G.~Niz, A.~Padilla and P.~M.~Saffin,
  ``Strong coupling in Ho\v{r}ava gravity,''
  JHEP {\bf 0908}, 070 (2009)
  [arXiv:0905.2579 [hep-th]].
 
\bibitem{bunch2}
  M.~Li and Y.~Pang,
  ``A Trouble with Ho\v{r}ava--Lifshitz Gravity,''
  JHEP {\bf 0908}, 015 (2009)
  [arXiv:0905.2751 [hep-th]].

\bibitem{bunch3}
  A.~A.~Kocharyan,
  ``Is nonrelativistic gravity possible?,''
  Phys.\ Rev.\  D {\bf 80}, 024026 (2009)
  [arXiv:0905.4204 [hep-th]].

\bibitem{bunch4}
  D.~Blas, O.~Pujolas and S.~Sibiryakov,
  ``On the Extra Mode and Inconsistency of Ho\v{r}ava Gravity,''
  JHEP {\bf 0910}, 029 (2009)
  [arXiv:0906.3046 [hep-th]].
  
\bibitem{Kobakhidze:2009zr}
  A.~Kobakhidze,
  ``On the infrared limit of Horava's gravity with the global Hamiltonian
  constraint,''
  Phys.\ Rev.\  D {\bf 82}, 064011 (2010)
  [arXiv:0906.5401 [hep-th]].
 
\bibitem{bunch5}
  A.~Wang and R.~Maartens,
  ``Linear perturbations of cosmological models in the Ho\v{r}ava--Lifshitz theory
  of gravity without detailed balance,''
  Phys.\ Rev.\  D {\bf 81}, 024009 (2010)
  [arXiv:0907.1748 [hep-th]].
 
\bibitem{bunch6}
   N.~Afshordi,
  ``Cuscuton and low energy limit of Ho\v{r}ava--Lifshitz gravity,''
  Phys.\ Rev.\  D {\bf 80}, 081502(R) (2009)
  [arXiv:0907.5201 [hep-th]].
 
 \bibitem{bunch7}
 K.~Koyama and F.~Arroja,
  ``Pathological behaviour of the scalar graviton in Ho\v{r}ava--Lifshitz gravity,''
  JHEP {\bf 1003}, 061 (2010)
  [arXiv:0910.1998 [hep-th]].
  
  
\bibitem{bunch8}
  M.~Henneaux, A.~Kleinschmidt and G.~L.~Gomez,
  ``A dynamical inconsistency of Ho\v{r}ava gravity,''
  Phys.\ Rev.\  D {\bf 81}, 064002 (2010)
  [arXiv:0912.0399 [hep-th]].
 
 \bibitem{bunch9}
  D.~Blas, O.~Pujolas, S.~Sibiryakov,
  ``Models of non-relativistic quantum gravity: The Good, the bad and the healthy,'' JHEP {\bf 1104}, 018 (2011)
 [ arXiv:1007.3503 [hep-th]].
  
\bibitem{Wang:2010uga}
  A.~Wang and Q.~Wu,
  ``Stability of spin-0 graviton and strong coupling in Horava-Lifshitz theory
  of gravity,''
  Phys.\ Rev.\  D {\bf 83}, 044025 (2011)
  [arXiv:1009.0268 [hep-th]].



  

\bibitem{Horava:2010zj}
  P.~Ho\v{r}ava and C.~M.~Melby-Thompson,
  ``General Covariance in Quantum Gravity at a Lifshitz Point,''
  Phys.\ Rev.\  D {\bf 82}, 064027 (2010)
  [arXiv:1007.2410 [hep-th]].

\bibitem{Sotiriou:2010wn}
  T.~P.~Sotiriou,
  ``Horava-Lifshitz gravity: a status report,''
  J.\ Phys.\ Conf.\ Ser.\  {\bf 283}, 012034 (2011)
  [arXiv:1010.3218 [hep-th]].


   
 \bibitem{Blas:2009qj}
  D.~Blas, O.~Pujolas and S.~Sibiryakov,
  ``Consistent Extension Of Ho\v{r}ava Gravity,''
  Phys.\ Rev.\ Lett.\  {\bf 104}, 181302 (2010)
  [arXiv:0909.3525 [hep-th]].
    

\bibitem{Papazoglou:2009fj}
  A.~Papazoglou and T.~P.~Sotiriou,
  ``Strong coupling in extended Ho\v{r}ava--Lifshitz gravity,''
  Phys.\ Lett.\  B {\bf 685}, 197 (2010)
  [arXiv:0911.1299 [hep-th]].
  
  
\bibitem{Kimpton:2010xi}
  I.~Kimpton and A.~Padilla,
  ``Lessons from the decoupling limit of Ho\v{r}ava gravity,''
  JHEP {\bf 1007}, 014 (2010)
  [arXiv:1003.5666 [hep-th]].


\bibitem{Blas:2009ck}
  D.~Blas, O.~Pujolas and S.~Sibiryakov,
  ``Comment on `Strong coupling in extended Ho\v{r}ava--Lifshitz gravity',''
  Phys.\ Lett.\  B {\bf 688}, 350 (2010)
  [arXiv:0912.0550 [hep-th]].


\bibitem{Jacobson:2010mx}
  T.~Jacobson,
  ``Extended Ho\v{r}ava gravity and Einstein-aether theory,''
  Phys.\ Rev.\  D {\bf 81}, 101502 (2010)
  [Erratum-ibid.\  D {\bf 82}, 129901 (2010)]
  [arXiv:1001.4823 [hep-th]].

  
   
    \bibitem{Jacobson:2000xp}
  T.~Jacobson and D.~Mattingly,
  ``Gravity with a dynamical preferred frame,''
  Phys.\ Rev.\  D {\bf 64}, 024028 (2001)
  [arXiv:gr-qc/0007031].


 
\bibitem{Jacobson:2008aj}
  T.~Jacobson,
  ``Einstein-aether gravity: a status report,''
  PoS {\bf QG-PH}, 020 (2007)
  [arXiv:0801.1547 [gr-qc]].

  \bibitem{Horava:2008ih}
  P.~Ho\v{r}ava,
  ``Membranes at Quantum Criticality,''
  JHEP {\bf 0903}, 020 (2009)
  [arXiv:0812.4287 [hep-th]].

\bibitem{Horava:2011gd}
  P.~Horava,
  ``General Covariance in Gravity at a Lifshitz Point,''
  Class.\ Quant.\ Grav.\  {\bf 28}, 114012 (2011)
  [arXiv:1101.1081 [hep-th]].



\bibitem{Eling:2006xg}
  C.~Eling and T.~Jacobson,
  ``Two-dimensional gravity with a dynamical aether,''
  Phys.\ Rev.\  D {\bf 74}, 084027 (2006)
  [arXiv:gr-qc/0608052].

\bibitem{poincare} H.~Poincar\'e, ``Sur l'uniformisation des fonctions analytiques," Acta Mathematica {\bf 31}, 1 (1908).  

 \bibitem{Germani:2009yt}
  C.~Germani, A.~Kehagias and K.~Sfetsos,
  ``Relativistic Quantum Gravity at a Lifshitz Point,''
  JHEP {\bf 0909}, 060 (2009)
  [arXiv:0906.1201 [hep-th]].
  



\end{thebibliography}
\end{document}